\documentclass[runningheads]{llncs}
\usepackage{graphicx}
\usepackage{mathrsfs}
\usepackage{bm}
\usepackage{amsmath}
\usepackage{amssymb}
\usepackage[colorlinks,linkcolor=blue, urlcolor=black]{hyperref}
\usepackage{makecell}
\usepackage{multirow}

\begin{document}
\title{Auto-weighting for Breast Cancer Classification in Multimodal Ultrasound}
\titlerunning{Auto-weighting for Breast Cancer Classification in Multimodal Ultrasound}
\authorrunning{Jian Wang, Juzheng Miao et al.}
\author{Jian Wang\textsuperscript{1}\thanks{Jian Wang and Juzheng Miao contribute equally to this work.} \and Juzheng Miao\textsuperscript{2$\star$} \and Xin Yang\textsuperscript{1} \and Rui Li\textsuperscript{1} \and Guangquan Zhou\textsuperscript{2} \and \\ Yuhao Huang\textsuperscript{1} \and Zehui Lin\textsuperscript{1} \and Wufeng Xue\textsuperscript{1} \and  Xiaohong Jia\textsuperscript{3} \and Jianqiao Zhou\textsuperscript{3} \and Ruobing Huang\textsuperscript{1}\thanks{Corresponding author, email: ruobing.huang@szu.edu.cn.} \and Dong Ni\textsuperscript{1}}
% 1{Wang,Jian}
% 1{Miao,Juzheng}
% 2{Yang,Xin}
% 3{Li,Rui}
% 4{Zhou,Guangquan}
% 5{Huang,yuhao}
% 6{Lin,Zehui}
% 7{Xue,Wufeng}
% 8{Jia,Xiaohong}
% 9{Zhou,Jianqiao}
% 10{Huang,Ruobing}
% 10{Ni,Dong}

\institute{\textsuperscript{$1$} Medical UltraSound Image Computing (MUSIC) Lab,  School of Biomedical Engineering, Shenzhen University, China\\
\textsuperscript{$2$}School of Biological Sciences and Medical Engineering, Southeast University, China\\
\textsuperscript{$3$}Department of Ultrasound Medicine, Ruijin Hospital, School of Medicine, Shanghai Jiaotong University, China }
%textsuperscript{$1$} National-Regional Key Technology Engineering Laboratory for Medical\\ Ultrasound, Guangdong Key Laboratory for Biomedical Measurements and \\Ultrasound Imaging, School of Biomedical Engineering, Health Science Center, \\Shenzhen University, Shenzhen, China\\
%\renewcommand\Authands{ and }

%
%\footnote{Corresponding Author: Ruobing}

%\institute{$\circ$ School of Biomedical Engineering, Shenzhen University\\$\dagger$Department of Ultrasound Medicine, Shanghai Ruijin Hospital;}
%
\maketitle              % typeset the header of the contribution
\begin{abstract}
Breast cancer is the most common invasive cancer in women. Besides the primary B-mode ultrasound screening, sonographers have explored the inclusion of Doppler, strain and shear-wave elasticity imaging to advance the diagnosis. However, recognizing useful patterns in all types of images and weighing up the significance of each modality can elude less-experienced clinicians. In this paper, we explore, for the first time, an automatic way to combine the four types of ultrasonography to discriminate between benign and malignant breast nodules. A novel multimodal network is proposed, along with promising learnability and simplicity to improve classification accuracy.  The key is using a weight-sharing strategy to encourage interactions between modalities and adopting an additional cross-modalities objective to integrate global information. In contrast to hardcoding the weights of each modality in the model, we embed it in a Reinforcement Learning framework to learn this weighting in an end-to-end manner. Thus the model is trained to seek the optimal multimodal combination without handcrafted heuristics. The proposed framework is evaluated on a dataset containing 1616 sets of multimodal images. Results showed that the model scored a high classification accuracy of 95.4\%, which indicates the efficiency of the proposed method.

\end{abstract}

\section{Introduction}
\begin{figure}[t]
	\centering
	\includegraphics[width=0.78\linewidth]{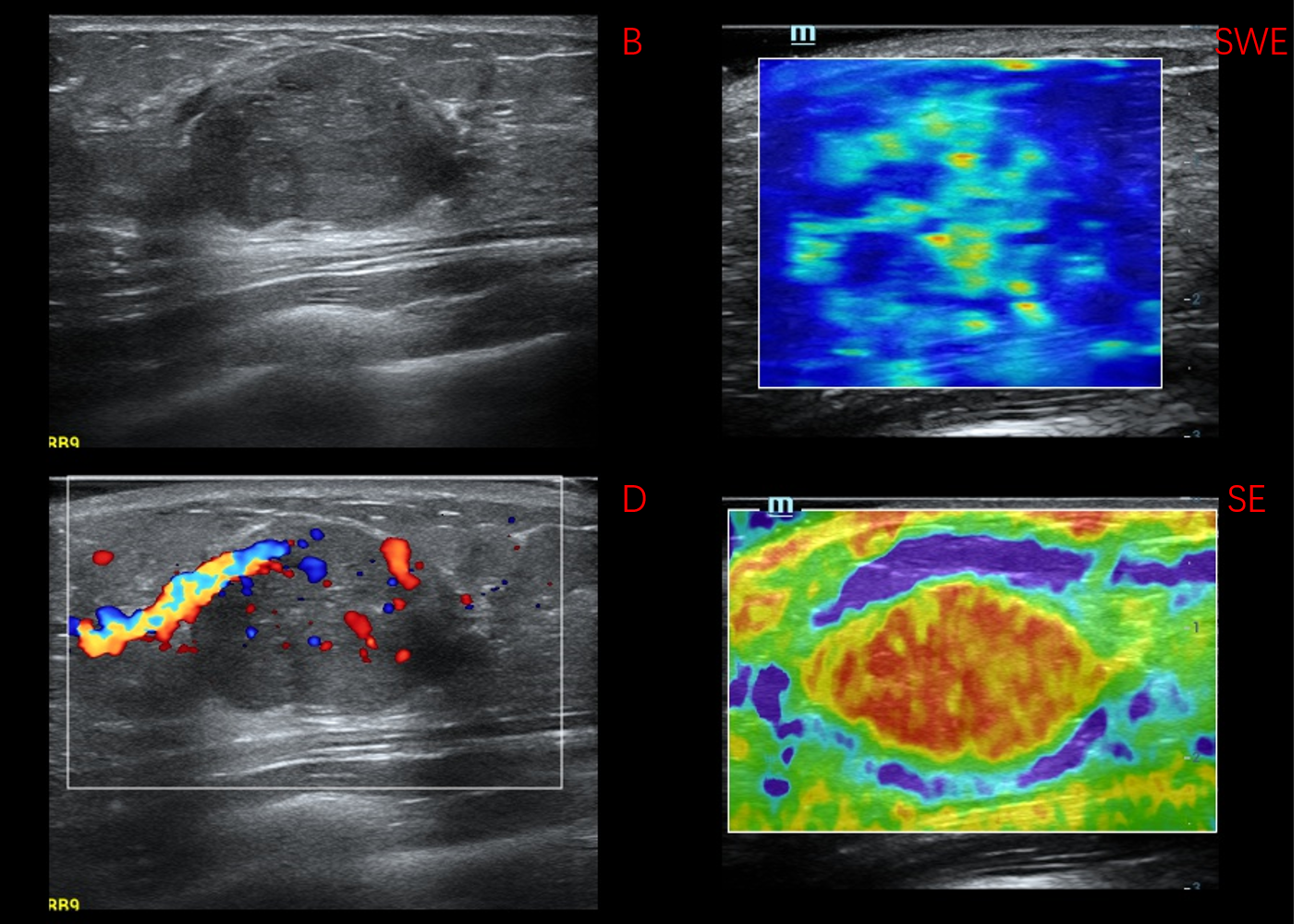}
	\caption{Multimodal ultrasound images (top to down). \emph{Left column}: B-mode and color doppler images. \emph{Right column}: Shear wave and strain elastography images.}
	\label{fig:intro}
\end{figure}

Breast cancer is one of the leading causes of cancer death in women~\cite{WHOorg2019}. Early and accurate diagnosis is crucial for better prognosis and the improvement of survival rate~\cite{ciatto1989t}. Mammography is the most common tool for breast cancer screening, while it could be less sensitive for patients with dense breasts~\cite{murtaza2019deep}, which might cause unnecessary biopsies or missed diagnosis~\cite{jesneck2007breast}. Being radiation-free, non-invasive, and relatively cheap, Ultrasound (US) has shown great potential in breast cancer diagnosis. To assist the most widely adopted B-mode ultrasound, researchers have adopted Color Doppler, Share Wave Elastography (SWE) and Strain Elastography (SE) images to improve the diagnosis accuracy (see Fig.~\ref{fig:intro}). As they provide valuable supplementary information such as vascularity and tissue stiffness. However, accurate diagnosis using rich multimodal images is extremely challenging and highly depends on the operators' experience.

Computer-Aided Diagnosis (CAD) methods using single-modality breast images have been developed. In \cite{liu2018integrate}, a multi-task learning (MTL) framework was proposed to classify breast B-mode images and predict tumor malignancy. Zhang et al. \cite{zhang2016deep} classified the SWE images into benign or malignant using a deep belief network. Recently, CAD methods of multimodal breast images have been studied. Sultan et al. \cite{sultan2018machine} employed traditional regression methods to differentiate breast lesion in the manually cropped B-mode and Doppler images. Zhang et al. \cite{zhang2019dual} further proposed a two-stage method to segment tumors from B-mode and SWE images and classify them into benign and malignant by a deep polynomial network. The above CAD methods only consider two modalities and can be promoted greatly by making use of state-of-the-art multimodal deep learning techniques. Good multimodal representations are crucial for the performance of machine learning models. They are categorized into joint and coordinated representations in \cite{baltruvsaitis2018multimodal}. While the former project different modalities into the same representation space, the latter enforces the representations of different modalities to be more complementary. Ge et al. \cite{ge2017skin} explored three representations based on an MTL framework, namely sole-net, share-net, and triple-net, for skin cancer diagnosis. The sole-net and share-net are joint and coordinated representations, respectively and the triple-net achieved best diagnosis accuracy because it made use of the advantages of both representations. Despite showing promising results,  existing methods usually treat different modalities equally, neglecting the fact that some of them should contribute more to the final decision-making.  How to assign proper weight to each of the modalities remains an open question.

In this paper, we propose a novel multi-modality classification framework, termed as Auto-weighting Multimodal (AWMM), to address the challenging problem in breast cancer diagnosis. Our contribution is threefold: 1) jointly employ all four types of ultrasonography, namely B-mode, Doppler, SWE and SE for the first time, which are used clinically for breast nodule identification. 2) we cast the task as a multi-task learning problem and adopt a weight-sharing mechanism to avoid overfitting with a mixed loss to integrate extracted representations; 3) to automatically obtain the optimal weighting between modalities, we embed the model into a Reinforcement Learning (RL) framework to learn the weighting in an end-to-end fashion. Experiment results showed that the trained model is able to process rich multimodal information and identify the breast nodules accurately.

\section{Methodology}
\begin{figure}[t]
	\centering
	\includegraphics[width=0.85\linewidth]{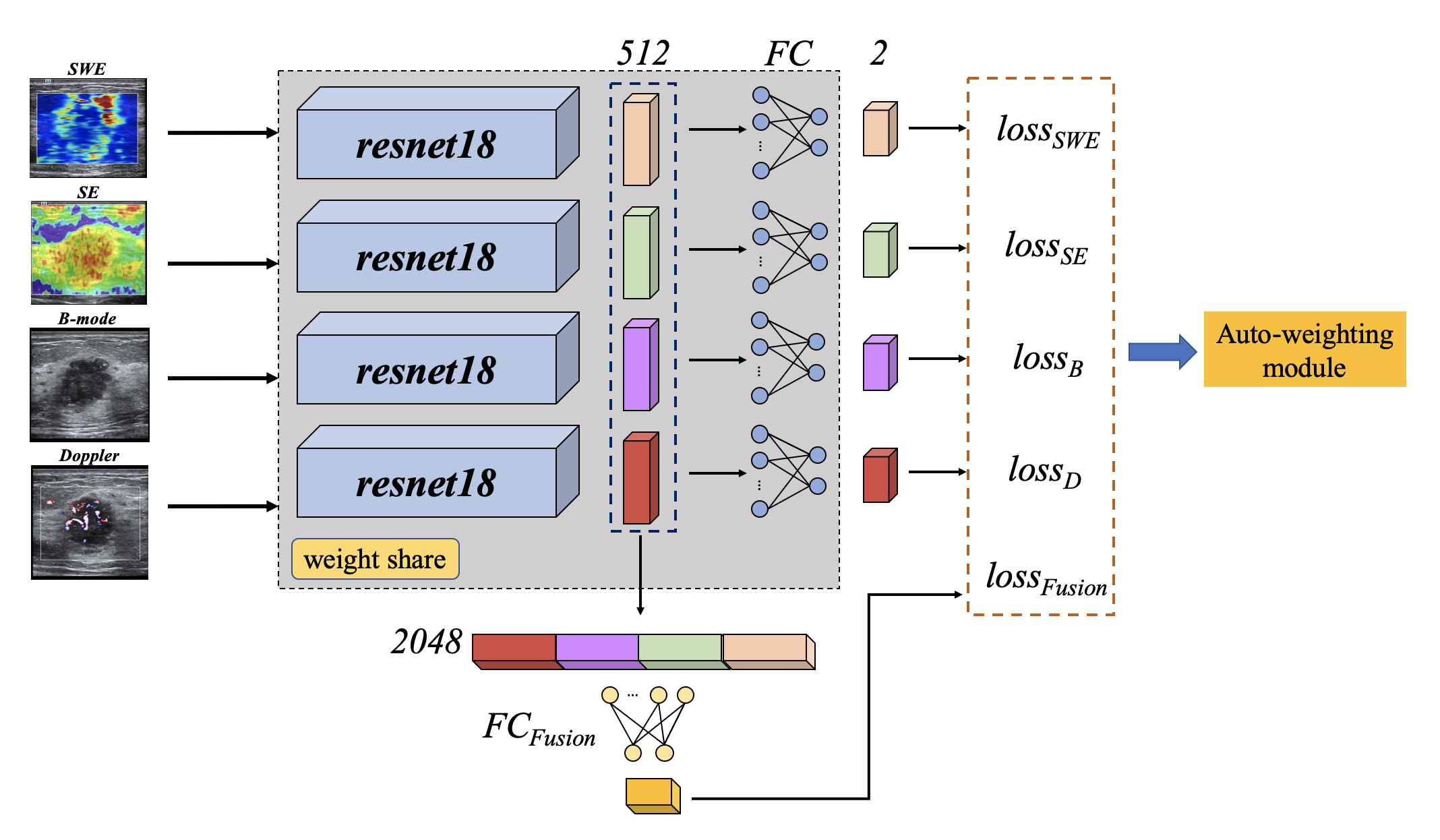}
	\caption{Schematic of the proposed multimodal learning model. It consists of four streams, each corresponds to one investigated modality (B-mode, Doppler, SE, SWE). A fusion loss (yellow cube) that merges all branches and other four losses are combined together.}
	\label{fig:schematic}
\end{figure}

In order to better mimic experts behaviour in diagnosing, we explore the advantages of linking information from four types of ultrasound imaging modalities through a multi-branch neural network model. The bespoke model is equipped with a weight-sharing mechanism and mixed fusion losses. The model is trained under an RL framework to achieve automatic branch (modality) weighting. As shown in Figure. \ref{fig:schematic}, the proposed model replicates ResNet-18 base model four times to construct a weight sharing MTL model to extract multimodal representations while suppress overfitting. Besides independent loss for each stream, an additional fusion loss is added to encourage competition. As opposed to manually select the weight of each branch, the model leverages an RL based framework to learn the optimal weighting between different modalities. This design can exempt the hand-crafted heuristics to weight different task as common multi-modal approaches. More importantly, it allows progressively interaction across modalities that contain different information and should contribute differently to the final prediction. Next, we explain the details of the framework.

\subsection{Learning Multimodal Representation}
Representing data in a format that a computational model can work with has always been a challenge in machine learning. It is especially true for multimodal models, which need to process and relate information from heterogeneous sources. Inspired by Share-Net \cite{ge2017skin}, the proposed model builds on four identical networks with shared-weights, each of which is able to make its own prediction. We adopt the Resnet-18 as the backbone that has been widely utilized in image recognition \cite{he2016deep}. We argue that this architecture is less prone to overfitting since the network is forced to extract features that are functional across different modalities while reducing the number of learnable parameters. On the other hand, each stream has its own objective which enables more flexibility in learning and might contribute to accuracy improvement. At the output level, a fully connected layer (${FC_{AWMM}}$) with a size of $2048\times2$ is added to project multimodal features into the shared space (see Fig.\ref{fig:schematic}). Then, this joint representation is used to compute a fusion loss as an additional objective.  It poses extra regularization that can help generalization (similar finding reported in \cite{vielzeuf2018centralnet}). Therefore, the proposed multimodal learning framework is further advanced at the output level, where both complementary and joint information can be learned using the combination of these loss functions. The model parameters are optimized by applying a stochastic gradient descent on the global loss, defined as:
$\sigma_{i}^{t,j}$

\begin{equation}
{Loss_{G}} = {\sum\limits {\alpha_{mod}}{Loss_{mod}}}
\end{equation}
, where $mod \in {SWE,SE,B,D,Fusion}$. ${Loss_{Fusion}}$ is the loss computed from the joint representation, and $Loss_{mod}$, $mod \in {SWE,SE,B,D}$ is the loss computed only using the $mod$ modality. The weights, $\alpha_{mod}$, are the coefficients to setup the trade-off between each modality and the cross-modality learning, which satisfy ${\sum\limits{\alpha_{mod}}}=1$. Our AWMM model allows the back-propagation of both modality-specific and universal signals for parameter update, thus helps to coordinate the learning process and further suppressing the overfitting.

\subsection{Auto-weighting modality using Reinforcement Learning Framework}

Similar to the behavior of clinical experts in making a diagnosis, an ideal model should make predictions using all the available information but puts emphasis on one or two of the imaging modalities. This could be incorporated into a multi-task model as weights associated with different tasks. It is impractical to define these weights manually, as even the most experienced experts have difficulty deciding if B-mode mode is 1.2 or 0.8 times more important than the elastic images.  Inspired by the work \cite{li2019lfs}, we enable the model to learn the optimal set of weighting using an RL framework in an end-to-end manner, avoiding the introduction of subjective bias.

In the classic setting of RL, an agent in its current state $\bm{S}$, interacts with the environments $\bm{E}$ by making successive actions ${\bm{\alpha}}\in{\bm{A}}$ that maximize the expectation of reward. In this study,  the environments $\bm{E}$ is the ultrasound images of different modalities. The state $\bm{S}$ is the set of weights, each of its elements corresponds to the weight of each task. These weights, represented by $\{{\alpha}_{mod}\}$, can be calculated by applying softmax to five real-valued agent output $\{{\beta}_{mod}\}:{\alpha}_{mod}=\frac{exp({\beta}_{mod})}{\sum{exp({\beta}_{mod})}}$. As the agent interacts with $\bm{E}$ to maximize the reward, the system can continuously adjust the weighting between different tasks. The action space $\{\Delta\beta_{mod}\}$ is defined as [-0.2,0,+0.2]. Each valid action gets a scalar reward $\bm{R}$ calculated based on the classification accuracy, which indicates whether the agent is moving towards the preferred target.

\begin{figure}[t]
	\centering
	\includegraphics[width=0.85\linewidth]{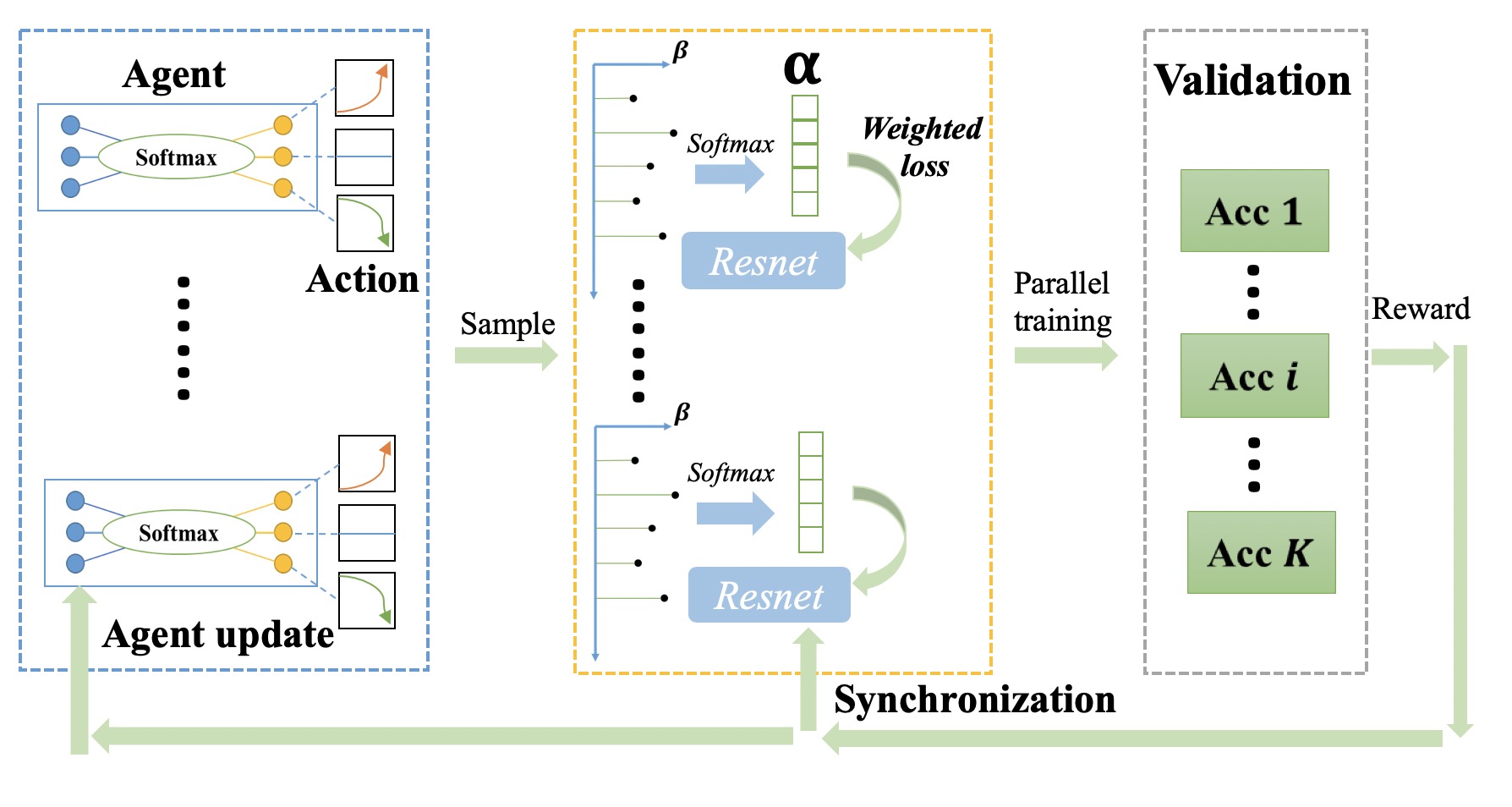}
	\caption{Learning process of the proposed framework.}
	\label{fig:pipeline}
\end{figure}

The learning process is a standard bi-level optimization problem, in which we aim to maximize the reward w.r.t. the controller parameters $\{\bm{\theta}\}$, as well as minimizing the loss of the network w.r.t. model parameters $\{\bm{\omega}\}$. $\{\bm{\theta}\}$ and $\{\bm{\omega}\}$ are updated alternately. As illustrated in Figure.\ref{fig:pipeline}, at the inner level, we sample $K$ times from the parameter space to generate $K$ different sets of task weights ${\{\alpha^{t,1:k}_{mod}\}}$ for $K$ identical networks (e.g. the whole model shown in Fig.\ref{fig:schematic}). These networks are then trained independently using the training set for one epoch to update their network parameters $\omega^k$, respectively. In the outer level of the optimization, the model having the maximum validation accuracy is selected to update $\{\bm{\theta}_{i}\}$ for each controller. Each controller corresponds to one modality and is independent to each other. They are updated iteratively using the REINFORCE rule \cite{williams1992simple}. In specific, the $i\rm{th}$ controller is updated following:
\begin{equation}
	\begin{split}
\bm{\Theta}_i^{t+t'}&=\bm{\Theta}_i^t+\eta\frac{1}{K}\sum_{j=1}^KR_j\cdot\nabla_{\bm{\theta}}log\left(g\left(\bm{\pmb{\alpha}}^{t,j}\right)\right) \\
&=\bm{\Theta}_i^t+\eta\frac{1}{K}\sum_{j=1}^KR_j\cdot\nabla_{\bm{\theta}}\sum log\left(p\left(\bm{\alpha}_{mod}^{t,j}\right)\right)
	\end{split}
\end{equation}
, where ${\eta}$ is the learning rate, and $g\left(\bm{\pmb{\alpha}}^{t,j}\right)=\prod p\left(\alpha_{mod}^{t,j}\right)$ is the joint probability distribution of four single-modality losses and the fusion loss. The reward signal, $R_j$, is calculated based on the validation accuracy of the corresponding $j\rm{th}$ network. After training the over-parameterized network, an optimal set of the parameters of the multimodal CNN model and desirable weights between tasks are simultaneously obtained.

\section{Materials and Experiments}
We validate our solution on a dataset of 1616 sets of multimodal ultrasound images of breast nodules. Approved by the local Institutional Review Board, the images were acquired from 835 different patients. Each set has a total of four images (each of B-mode, Color Doppler, SWE, and SE modality) collected from the same patient. We randomly split the dataset into 60$\%$ training, 20$\%$ validation, and 20$\%$ test at the patient level. All images are resized to $544 \times 320$. Biopsies were carried out for each subject to classified whether the nodule is benign or malignant, and the result is used as the ground truth in this experiment.

To validate the effectiveness of the proposed model, comparison experiments were carried out using single-modality models and various popular multimodal approaches, including Share-Net \cite{ge2017skin}, Centralnet \cite{vielzeuf2018centralnet}, and Voting Schemes\cite{morvant2014majority}. We termed our Manual-weighting MultiModal framework with or without sharing network parameters as MWMM-SN, MWMM respectively and its Auto-weighting version as AWMM-SN. All experiments are implemented in PyTorch with a GeForce RTX 2080Ti GPU.

Different augmentation strategies were applied, including scaling, rotation (up to ${30^\circ}$), flipping, and mixup. Weights pre-trained from ImageNet were used for initializing. We use the Adam optimizer with a learning rate of 1e-4. For the MWMM framework, the weights of each task's loss are equal ($\alpha_{mod} = 0.2$). Meanwhile, we set the sample number $K = 10$ and all controllers value $\beta_{mod}$ to 1 for employing the RL based weight tuning framework. The controller parameters {$\Theta$} is trained using Adam optimizer with learning rate 1e-3.

\section{Results and Discussion}
\begin{table}
\caption{Performance evaluation of single-modality and multi-modality methods. Sensitivity (SEN), specificity (SPE), overall accuracies (ACC), precision with the cut-off value of 0.5 (PRE), and F-score metric (F1-score) are used.}\label{Classification results}
\begin{center}
\scriptsize
\begin{tabular}{c|c|c|c|c|c|c}
\hline
\hline
& Methods & ACC(\%) & SEN(\%) & SPE(\%) & PRE(\%) & F1-score(\%) \\
\hline
\multirow{4}*{Single-modality} &SWE &91.46&91.03&91.86&91.03 & 91.03 \\
\cline{2-7}
&SE &89.02&88.46&89.53&88.46&88.46\\
\cline{2-7}
&B-mode &86.28&83.97&88.37&86.75&85.34\\
\cline{2-7}
&Doppler &86.59&82.05&90.70&88.89&85.33\\
\hline

\hline
\multirow{6}*{Multi-modality}
&Voting &91.16&91.03&91.28& 90.45&90.74 \\
\cline{2-7}
&Centralnet &92.99& 91.67& 94.19&93.46&92.56\\
\cline{2-7}
&Sharenet &93.29 &\textcolor{blue}{96.79} &90.12& 89.88&93.21\\
\cline{2-7}
&MWMM &92.99&92.95&93.02&92.36&92.65\\
\cline{2-7}
&MWMM-SN  &93.60&92.95&94.19&93.55&93.25\\
\cline{2-7}
&AWMM-SN &\textcolor{blue}{95.43} &96.15 &\textcolor{blue}{94.77} &\textcolor{blue}{94.34} &\textcolor{blue}{95.24}\\
\hline
\hline
\end{tabular}
\label{tab:results}
\end{center}
\end{table}

We first compare the performance of single modality models, each of which only had access to one of the four investigated modalities. As shown in Table.~\ref{tab:results}, the SWE model achieved the best performance in Accuracy(ACC )=91.46\%, Sensitivity(SEN)=91.03\%, Specificity(SPE)=91.86\%, Precision(PRE)= 91.03\%, and F1-score=91.03\%. It indicated that the stiffness of nodule tissue might be a strong indicator of its malignancy. On the contrary, the vascularity of nodule seemed to be less important, as the single modality model using only Doppler images achieved relatively lower accuracy (row 4 in Tab.\ref{tab:results}). This is in accord with our clinical collaborators' finding in the real world situation, as a malignant breast nodule can have either rich or scarce blood flow.

Compared with single modality models, multi-modalities models showed better performance overall, whichever combination strategies they used. This proves that there is indeed some complementary information hidden in different types of ultrasonography, and their combination could lead to a more accurate diagnosis. Furthermore, it can be observed that both the proposed models and the Share-Net outperformed the CentralNet and the Voting Schemes approaches. This might result from the weight sharing strategy which forces the model to extract more universal features and also helps to reduce the overfitting by introducing fewer parameters. The last two rows in Tab.\ref{tab:results} showed that the necessity of the RL-based auto-weighting process as AWMM-SN (the proposed method) outperformed the MWMM-SN method, which weights each modality equally. In specific, the automatically learned weights for each branch are: SWE: 0.15, SW:0.19, B-mode: 0.13, Doppler: 0.19, and Fusion: 0.34. This further validates the importance of incorporating all the modalities in decision-making and verifies our model design. Another interesting observation is that the SWE did not gain the highest weight as expected. We conjecture that there might exist certain competition between SWE and SW modalities that lead to this phenomenon. Note that our model does not require the presence of all four modalities to predict and still works well when the modalities are incomplete. The proposed model achieved Acc of 91.2\% (SWE), 86.0\% (SE), 84.8\% (B-mode), and 86.6\% (Doppler) when only one modality is available during the test. The key challenge here is to extract and combine multi-modal information efficiently.

%As in the real clinical setting, some of the modalities might not be available during inference. Thus, we test the proposed model

 \begin{figure}
 \centering
 \includegraphics[width=1.0\textwidth]{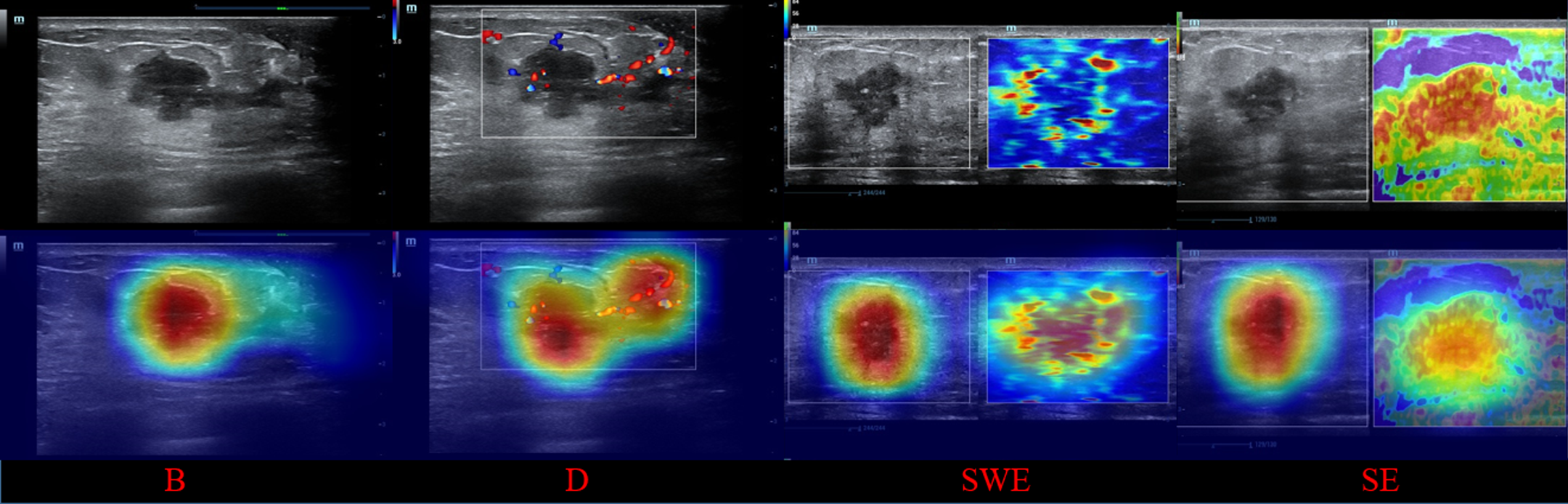}
 \caption{Grad-CAM Visualization of AWMM-SN for one sample. Note that the network focused on the nodule region despite no location information was given in training. }
 \label{Grad-CAM}
\end{figure}

\subsubsection{Ablation study}
 As an ablation study, we tested whether removing the weight sharing scheme would effect the model performance. As shown in Tab.~\ref{tab:results}, the MWMM model achieved better classification accuracy than that of the single-modality methods, but was inferior to the proposed method. This result indicates that the multimodal information is beneficial to accuracy improvement, while joint representation plays a vital role in the multimodal representations. %Meanwhile, the proposed auto-weighting scheme for various tasks adaptively determines the weight and makes our AWMM-SN model achieve the highest ACC, SEN, SPE, and F1-score.

We further employed Gradient-weighted Class Activation Mapping (Grad-CAM) to visualize the classification decisions of AWMM-SN model (Fig.~\ref{Grad-CAM}). It demonstrated that our model focused on the nodule regions during classification despite no location information was given for any of the modes. Note that no preprocessing was done to align the images, while the model automatically focused on the corresponding regions.  It should be noted that clinical experts usually rely on the shape prior in the B-mode image to guide the diagnosis of SWE or SE. We, therefore, keep the B-mode region in the SWE or SE modality.

\section{Conclusions}
In this paper, we proposed a novel multimodal framework that is able to process four types of ultrasound modalities for automatic classification of breast nodules. The model design encourages interactions between different modalities while allowing each contributes to the final prediction.  It is trained under a Reinforcement Learning framework to automatically find the optimal weighting across modalities to increase accuracy. As the design of this framework is general, it could be applied to other multimodal applications in the future.

\section{Acknowledgement}
This work was supported by National Key R\&D Program of China (No.2019YFC\\0118300); Shenzhen Peacock Plan (No. KQTD2016053112051497, KQJSCX2018\\0328095606003); Medical Scientific Research Foundation of Guangdong Province, China (No. B2018031); National Natural Science Foundation of China (Project No.NSFC61771130).

% ---- Bibliography ----
%
% BibTeX users should specify bibliography style 'splncs04'.
% References will then be sorted and formatted in the correct style.
%
\bibliographystyle{splncs04}
\bibliography{paper2994}

\end{document}